\documentclass[pss]{wiley2sp} 
\usepackage{amsmath}
\usepackage{bm}              
\usepackage{w-greek}

\def\efat{\mbox{\boldmath$\varepsilon$}}

\newcommand{\bk}{\mathbf{k}}
\newcommand{\bR}{\mathbf{R}}

\def\fd{f^\dagger}

\def\ket{\rangle}

\def\bran{\langle n|}

\def\phiAn{\phi_{An}^{\hfill}}

\begin{document}

\title{LDA+Slave-Boson approach to the correlated electronic structure 
of the metamagnetic bilayer ruthenate Sr$_3$Ru$_2$O$_7$}

\titlerunning{LDA+RISB for Sr$_3$Ru$_2$O$_7$}

\author{Christoph Piefke and Frank Lechermann\textsuperscript{\Ast}}

\authorrunning{C.~Piefke and F.~Lechermann}

\mail{e-mail
  \textsf{Frank.Lechermann@physnet.uni-hamburg.de}, Phone:
  +49-40-428387943, Fax: +49-40-428386798}

\institute{I. Institut f\"ur Theoretische Physik,
Universit\"at Hamburg, D-20355 Hamburg, Germany}

\received{XXXX, revised XXXX, accepted XXXX} 
\published{XXXX} 

\keywords{Strong correlation, electronic structure, metamagnetism,
density functional theory}

\abstract{%
%
%
%
\abstcol{%
The combination of the local-density approximation (LDA) with the rotationally
invariant slave-boson theory (RISB) is used to investigate the realistic correlated 
electronic structure of Sr$_3$Ru$_2$O$_7$. From Wannier-downfolding the 
low-energy band structure to a three-band model for the Ru($t_{2g}$) states, the
interacting problem is solved including intra- and inter-orbital Hubbard terms
as well as spin-flip and pair-hopping interactions. Therewith it is possible to 
obtain valuable insight into the orbital occupations, relevant local spin 
multiplets and the fermiology with increasing correlation strength.}
{Besides generic correlation-induced band-narrowing and -shifting, an 
intriguing quasiparticle structure close to the Fermi level is found in the 
neigborhood of the notorious $\gamma_2$ pocket in the Brillouin zone. Along
the $\Gamma$$-$$X$ direction in {\bf k}-space, that structure appears very 
sensitive to electronic self-energy effects. The subtle sensitivity, connected also
its manifest multi-orbital character, may put this very low-energy structure in 
context with the puzzling metamagnetic properties of the compound.}}
%
%
%

\maketitle   

\section{Introduction}
Strongly correlated electron systems are not only a fascinating research field from
the viewpoint of fundamental research, but also become more and more of 
technological relevance in the context of the demand for specific high-responsive 
behavior. The very subtle balance between the kinetic-energy term and the  
Coulomb repulsion among the electrons in these systems indeed often result in e.g.
absorbing magnetic, superconducting or thermoelectric properties. Concerning the
theoretical description on an atomistic level, the standard Kohn-Sham (KS) 
band-theory representation of density functional theory (DFT) for the solid state
is usually inappropriate for realistic materials with strong electronic 
correlations. For instance, the Mott-insulating state as an electron localization 
in real space amounts to a complete breakdown of the conventional band-structure 
concept, e.g., given by the local-density approximation (LDA) to DFT. However the
simple LDA+U extension for the strongly correlated regime is generally not 
well-suited for metals, since the correlated low-energy behavior close to the 
Fermi level $\varepsilon_{\rm F}$ cannot be captured within this static mean-field method.

Here we want to show that the combination of LDA with the slave-boson 
theory~\cite{col84,kot86} in its rotationally-invariant form~\cite{li89,lec07} 
(RISB) at saddle-point may be an efficient approach to model strongly correlated 
metals. This LDA+RISB scheme can account for many details of the quasiparticle (QP) 
physics close to $\varepsilon_{\rm F}$ of strongly interacting multi-orbital 
systems with modest effort compared to, e.g., numerically expensive 
Quantum-Monte-Carlo impurity computations within the framework of dynamical 
mean-field theory (DMFT).

In this work the electronic correlations effects in the complex Sr$_3$Ru$_2$O$_7$ 
compound shall be discussed within the LDA+RISB method. The puzzling ruthenate
has gathered much attention in recent years due to its metamagnetic
behavior at low temperatures, which may be related to a quantum critical 
point~\cite{gri01,geg06,bor06}. The paramagnetic Fermi-liquid displays one of the
highest electronic specific heat values among oxides and has definitely to be 
placed in the strongly correlated regime. Because of its very delicate low-energy
physics, the slave-boson technique, focussing on the strongly renormalized QP
part, is an adequate framework for investigations beyond standard LDA approaches. 

\section{The Sr$_3$Ru$_2$O$_7$ compound}
\begin{figure}[t]%
\begin{center}
\includegraphics*[width=8cm]{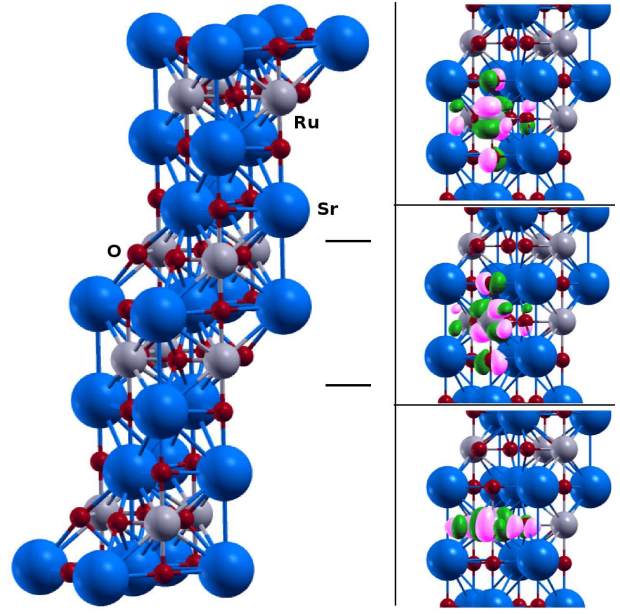}
\caption{Left: crystal structure of the orthorhombic $Bbcb$-Sr$_3$Ru$_2$O$_7$, where
the Ru bilayer part is indicated within the dark lines. Right from top to
bottom: $d_{yz}$, $d_{xz}$ and $d_{xy}$ effective Wannier function derived from 
the LDA band structure.
\label{pic:struc}}
\end{center}
\end{figure}

The $n$=2 case of the multilayer Ruddlesden-Popper strontium ruthenates
Sr$_{n+1}$Ru$_n$O$_{2n+3}$ serves as an interesting application of our
theoretical methodology. The perovskite end member SrRuO$_3$ ($n$=$\infty$) of 
this family is at low temperature $T$ a ferromagnetic metal, whereas the also
itinerant single-layered Sr$_2$RuO$_4$ compound is paramagnetic at ambient $T$
and becomes superconducting below $T_c$$\sim$4.2 K, with a widely believed triplet 
pairing~\cite{mac03}. The crystal structure of Sr$_3$Ru$_2$O$_7$~\cite{sha00} 
with lattice parameters $a$=$b$=5.5006{\AA} and $c$=20.725{\AA} is based on the 
orthorhombic space group $Bbcb$ ($\#$68) (see Fig.~\ref{pic:struc}) and does not 
show the fourfold symmetry of the simpler Sr$_2$RuO$_4$ structure. It consists of 
RuO$_2$ bilayers, whereby the RuO$_6$ octahedra are rotated by 6.8$^{\circ}$. The
unit cell exhibits four Ru ions, all equivalent by symmetry.

Many experiments show that the Sr$_3$Ru$_2$O$_7$ compound is located just at the
paramagnetic-to-ferromagnetic transition boundary. It is still paramagnetic in 
zero magnetic field, but positioned very close to the ferromagnetic 
instability~\cite{ike00}. The system shows strongly renormalized Fermi-liquid 
behavior with however a strong resistivity anisotropy 
$\rho_c/\rho_{ab}$$\sim$100~\cite{ike00}. With applied magnetic 
field $H$, the systems shows metamagnetic behavior, i.e., a very large 
$\partial M/\partial H$, around $H$=5.5(7.7)T for $H$$||$$ab$($c$)~\cite{perr01}.
Furthermore this metamagnetic region may be associated with being in the 
neigborhood of a quantum critial point that can approached via tuning the 
magnetic-field angle with the $ab$ plane~\cite{gri03}. Reachable within 
fields $H$$<$10 T, the physics of this metamagnetism is acting on a very low 
energy scale of the order of a few meV. Note also that the Fermi-liquid regime in 
vanishing field exists below 10-15 K, however can be driven to zero temperature 
with applied field~\cite{gri01}.

\section{Theoretical Approach}
In the present modeling we combine the LDA approach to DFT with the rotationally
invariant slave-boson theory at saddle-point. For the LDA part, we 
utilized an implementation~\cite{mbpp_code} of the highly-accurate mixed-basis 
pseudopotential (MBPP) technique~\cite{lou79}. This band-structure code employs 
normconserving pseudopotentials~\cite{van85} and an efficient mixed basis 
consisting of plane waves and a few localized orbitals. Scalar-relativistic
pseudopotentials and the LDA exchange-correlation functional after Perdew and
Wang~\cite{per92} were used in the actual computations.

Based on the LDA electronic structure a low-energy KS dispersion
$\efat^{\rm (KS)}_{\bk}$ for each wave vector $\bk$ within 
the $t_{2g}$ manifold of Sr$_3$Ru$_2$O$_7$ is constructed via maximally-localized
Wannier functions (MLWF)~\cite{mar97,sou01} for the four Ru ions in the 
orthorhombic unit cell. This effective single-particle $12\times 12$
Hamiltonian is then supplemented by onsite Coulomb interactions, resulting in
additional Hubbard, exchange, spin-flip and pair-hopping terms on each Ru ion $i$
with $t_{2g}$-orbital indices $m,m'$ in the unit cell $\alpha$. 
The interacting problem hence reads:
\begin{equation}
H=\sum_{\bk mm'\sigma}\varepsilon^{\rm (KS)}_{\bk mm'}\,
d^\dagger_{\bk m\sigma}d^{\hfill}_{\bk m'\sigma}+\sum_\alpha 
H^{\rm (loc)}_\alpha\quad,
\label{eq:fullham}
\end{equation}
with the local interacting part in the unit cell given by three-orbital generalized
Hubbard model on each Ru site, i.e.,
\begin{eqnarray}
H^{\rm (loc)}_\alpha&=&U\sum_{im} n_{im\uparrow}n_{im\downarrow}
+\frac 12 \sum \limits _{i,m \ne m',\sigma}
\Big\{U' \, n_{im \sigma} n_{im' \bar \sigma}\nonumber\\
&&+ U'' \,n_{im \sigma}n_{im' \sigma}+
J \, d^\dagger_{im \sigma} d^\dagger_{im' \bar\sigma} 
d^{\hfill}_{im \bar \sigma} d^{\hfill}_{im' \sigma}\nonumber\\
&&+\left. J_{\mathrm C} \,d^\dagger_{im \sigma} d^\dagger_{im \bar \sigma}
 d^{\hfill}_{im' \bar \sigma} d^{\hfill}_{im' \sigma}\right\}
\label{eq:locham}
\end{eqnarray}
with $n$=$d^\dagger d$. The first term in eq. (\ref{eq:locham}) marks the 
intra-orbital Coulomb interaction with Hubbard $U$, the second term the 
inter-orbital Hund's rule corrected interaction with $U'$=$U$$-$$J$ and 
$U''$=$U$$-$$2J$ for unequal and equal spin orientiation 
$\sigma$=$\uparrow,\downarrow$, respectively. Finally, the third term accounts for 
spin-flip and the fourth term for pair-hopping processes, were we used 
$J_{\rm C}$=$J$ for the real MLWFs. 

For the solution of the problem posed by eq. (\ref{eq:fullham}) the RISB method
was employed, where the electron operator 
$d^{\hfill}_{im\sigma}$$\equiv$$d^{\hfill}_{\mu\sigma}$ is represented as 
$\underline{d}_{\mu\sigma}$=
$\hat{R}[\phi]^{\sigma\sigma'}_{\mu\mu'}f_{\mu'\sigma'}$. Here $\hat{R}$ is a 
non-diagonal transformation operator that relates the physical 
operator to the QP operator $f_{\mu\sigma}$ and is written in terms of the 
slave bosons $\{\phi_{An}\}$. In this generalized slave-boson theory $\phi$ 
carries the index $A$ for the physical-electron state and $n$ for the QP Fock 
state. In the end, the general idea is to rewrite eq. (\ref{eq:locham}) solely
in terms of slave-boson operators which allows to integrate out the QP part.
The operator decomposition into QP and interacting slave-boson part introduces
two constraints, namely
\begin{eqnarray}
\sum_{An}\phi^\dagger_{An}\phiAn&=&1\quad,\label{sbcon1} \\ 
\sum_{Ann'} \phi^\dagger_{An'}\phiAn\bran\fd_{\mu\sigma} 
f_{\mu'\sigma'}^{\hfill}|n'\ket&=&f_{\mu\sigma}^{\dagger}\,f_{\mu'\sigma'}^{\hfill}
\label{sbcon2}\quad,\end{eqnarray}
whereby eq. (\ref{sbcon1}) normalizes the total boson weight to unity and 
eq. (\ref{sbcon2}) ensures that QP and boson contents match at every site $i$.
This selection of the physical states is imposed through a set of Lagrange 
multipliers $\{\lambda\}$. In the mean-field version at saddle-point these
constraints hold on average, with the bosons condensed to $c$
numbers. The numerical effort amounts to the solution of the saddle-point 
equations for $\{\phi\},\{\lambda\}$ (see Ref.~\cite{lec07} for further
details). In the calculations we also used the set of local Fock states $\{n\}$ for
the set of atomic states $\{A\}$ on each Ru ion. It is important to note in this
context that in the present work no inter-site slave bosons were introduced. Thus 
at the saddle-point, the RISB physical self-energy of the $d$ electrons is given 
by~\cite{lec07}
\begin{equation}
\mathbf{\Sigma}_d^{i}(\omega)=\omega\left(1-\mathbf{Z}^{-1}\right)\,
+[\mathbf{R}^\dagger]^{-1}\mathbf{\Lambda}\mathbf{R}^{-1}-\efat^{(0)}\,\,,
\label{eq:Sigma_physical}
\end{equation}
diagonal in the Ru sites. Here ${\bf Z}$=$\bR\bR^{\dagger}$ is the 
QP-weight matrix and ${\bf\Lambda}$ is
the lagrange-multiplier matrix. The quantity $\efat^{(0)}$ denotes a possible
one-body term in the k-summed KS-Hamiltonian, which should not appear in the
QP part of the framework. Thus $\Sigma_d$ contains a term linear in frequency 
and a static part. Though more approximative than more involved methods like, 
e.g., quantum Monte-Carlo, which may handle the full frequency dependence, this 
proves sufficient in many cases to describe the QP physics at low-energy.
It also provides additional insight into local excitations within a static 
self-energy approximation. In this regard, the slave-boson amplitudes yield direct 
access to the occupation of local multiplets in the metallic state. 
The RISB mean-field approach may therefore be seen as meaningful simplified
approximation to correlated metals beyond LDA. Note that it is superior to LDA+U
in this regime, since the latter is designed for Mott-insulating systems.

Our application of LDA+RISB to Sr$_3$Ru$_2$O$_7$ considers no spin-orbit-coupling 
(SOC) effects. Albeit SOC is believed to be important for the
low-energy physics of this compound, the present study shall exhibit that many
interesting features of the correlated electronic structure may already be 
revealed without that coupling. Further studies including the SOC with the 
obviously well-tailored RISB method will be discussed in a future work.

\section{Results and discussion}

\subsection{LDA electronic structure}

The Ru ions in the Sr$_3$Ru$_2$O$_7$ compound are formally in the $4+$ state, i.e.
the $4d$ shell of the transition metal is filled with four electrons. Since the
system is known to be in the low-spin state, all these electrons are located in
the $t_{2g}$ manifold. Many issues of the electronic structure on the LDA level
have already been discussed in previous studies~\cite{has97,sin01,tam08}, albeit
the very low-energy regime has not been addressed theoretically in great detail.

\begin{figure}[b]%
\begin{center}
\includegraphics*[width=4cm]{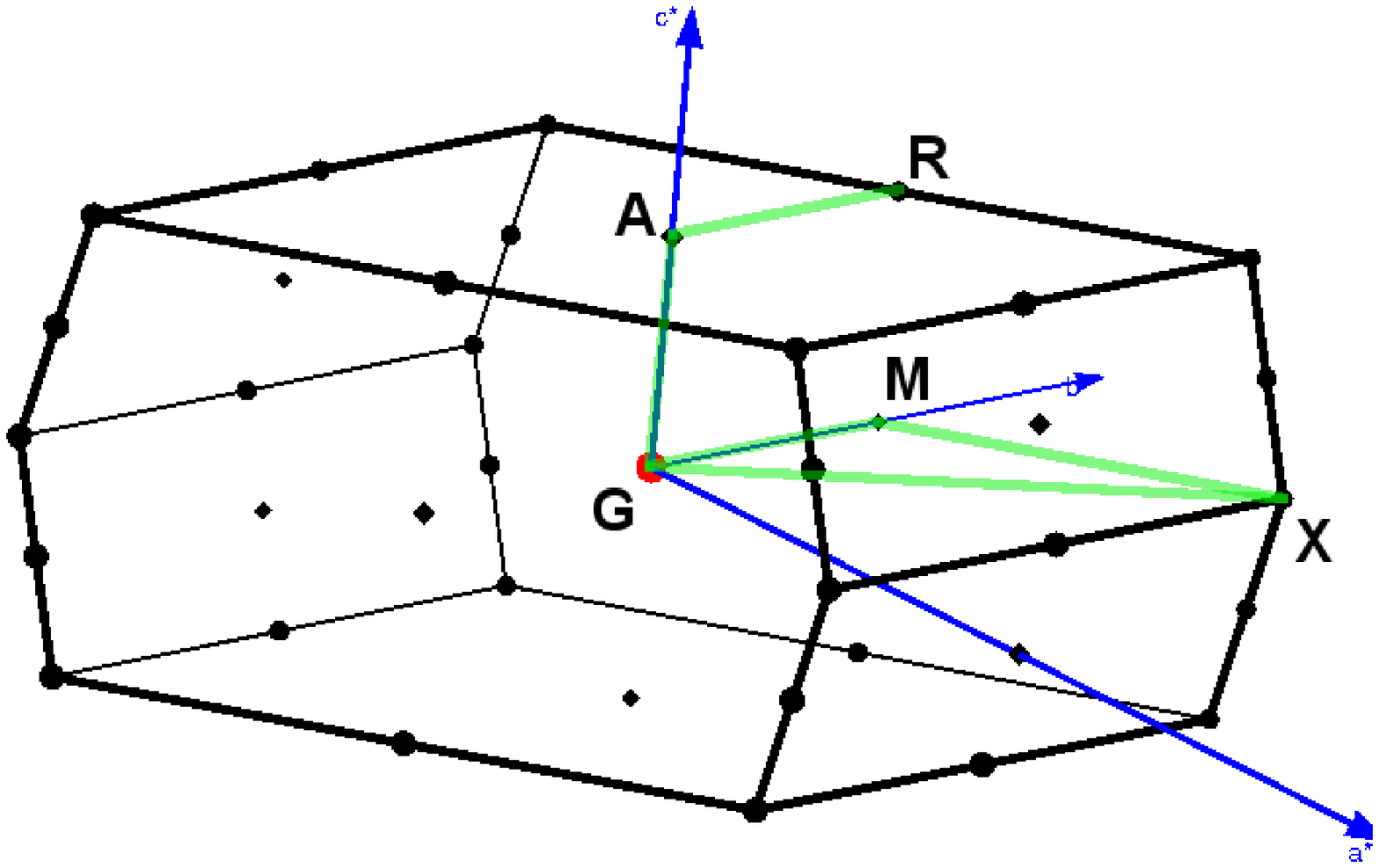}
\includegraphics*[width=7.5cm]{finalbands.eps}
\caption{Brillouin zone with selected high-symmetry points (top). The green line
marks the chosen path for the LDA band structure (bottom) including the downfolded 
effective Wannier $t_{2g}$ low-energy bands (cyan/grey lines).
\label{pic:ldabands}}
\end{center}
\end{figure}

Figure~\ref{pic:ldabands} shows the LDA band structure close to the Fermi level,
consisting mainly of an Ru$(t_{2g})$ dominated 12-band manifold. The dispersion
is strongly two-dimensional, although still some bands display significant
variation along $\Gamma$$-$$A$ in the first Brilluoin zone (BZ).
There are several rather flat bands close to $\varepsilon_{\rm F}$, especially in 
the neighborhood of the $X$ point, as also observed in recent angle-resolved 
photoemission (ARPES) experiments~\cite{tam08,lee09}. Additionally shown in 
Fig.~\ref{pic:ldabands} are the effective bands from the MLWF construction, which 
are in very good agreement with the original LDA bands close to zero energy. Due 
to the strong hybridization of the $d_{xy}$ orbital with oxygen states at energies 
below -2 eV, the effective Wannier manifold ranges from -2.5 eV to 0.6 eV, i.e.,
displays a bandwidth of about 3.1 eV.

\begin{figure}[t]%
\begin{center}
\includegraphics*[width=7.5cm]{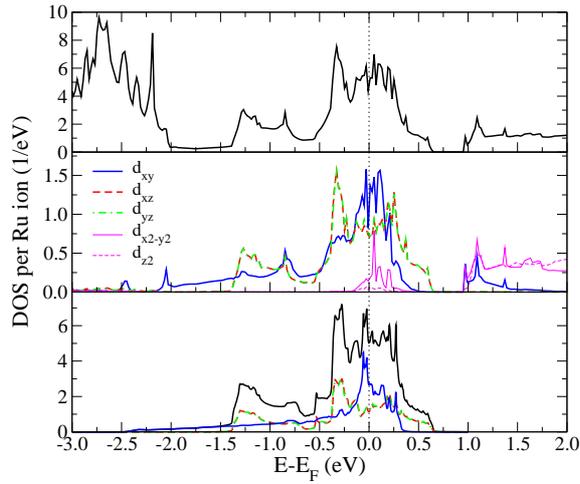}
\caption{LDA density of states normalized to one Ru ion. Top: total DOS, middle:
local Ru-$d$ DOS within $r_{\rm Ru}$=2.0 a.u.. Bottom: effective Wannier $t_{2g}$
low-energy DOS. Note that the $t_{2g}$ Wannier functions are tailored to match the
LDA bands and are thus to some extent build on linear combinations of the original 
$d$ states shown in the middle part.
\label{pic:ldados}}
\end{center}
\end{figure}

\begin{figure}[t]%
\begin{center}
\includegraphics*[width=6cm]{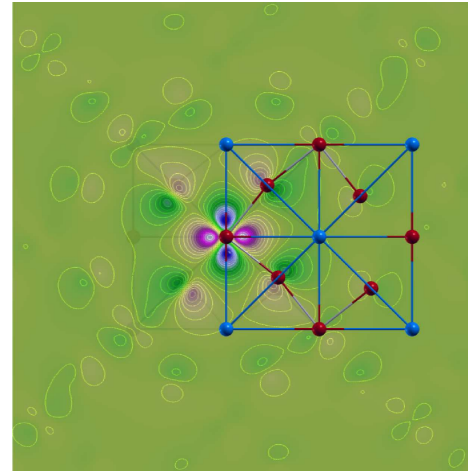}
\caption{Isolines of the $d_{xy}$-like Wannier function. The $x$- and $y$-axis
correspond to the diagonals of the plot. The Ru ion associated to the plotted
MLWF resides below the (red) oxygen ion.
\label{pic:isodxy}}
\end{center}
\end{figure}

More details of the LDA spectral behavior of the Ru$(4d)$ states may be
extracted from the density-of-states (DOS) plot in Fig.~\ref{pic:ldados}. It is
first observable that the Fermi level resides actually between prominent peaks
within the total DOS of the system. Projecting the local DOS onto the standard cubic
$d$ harmonics along the cartesian axes renders the dominant behavior of the 
$t_{2g}$ manifold at low energy obvious. Thereby the in-plane $d_{xy}$ orbital
has more substantial structure close to $E_F$ than the nearly degenerate
$d_{xz},d_{yz}$ orbitals. Interestingly, the $e_g$ orbital $d_{x^2-y^2}$ has
also some appreciable weight in that energy regime, with important overall
clear signatures of hybridization with the $d_{xy}$ orbital. The effective Wannier
DOS reproduces well the low-energy regime, but note that the obtained $t_{2g}$-only
Wannier orbitals are now not aligned along the cartesian axes anymore, but show
some tilting in line with the orthorhombic distortions of the $Bbcb$ structure.
Furthermore especially the effective $d_{xy}$-like orbital is now ``dressed'' with
the $d_{x^2-y^2}$ hybridization just mentioned. This is indeed visible in the
contour plot of the former Wannier orbital exhibited in Fig.~\ref{pic:isodxy},
which also shows the slight tilting in the $xy$ plane parallel to the RuO$_2$ 
bilayer. Note that from the underlying MLWF construction the on-site energy of the 
$d_{xy}$-like orbital is higher than for the $d_{xz,yz}$-like by about 115 meV. 

\subsection{LDA+RISB modeling}
In order to account for electronic on-site correlations beyond LDA we introduced
the local Hamiltonian~(\ref{eq:locham}) on each Ru site and solved the interacting
problem~(\ref{eq:fullham}) with the KS $t_{2g}$-like Wannier dispersions discussed 
in the last section. In the following we will refer to the latter by 
$d_{xz,yz,xy}$, albeit it is understood that the effective orbitals are tailored to 
$Bbcb$-Sr$_3$Ru$_2$O$_7$ through the present MLWF construction. The $d_{xz,yz}$
are only nearly degenerate, but in the discussion the orbitally averaged data
is shown, since the present orbital-resolved results differ only marginally. Note 
however that the calculations allowed for the full differences between all the 
treated local orbitals. 
\begin{figure}[b]%
\begin{center}
\includegraphics*[width=6.5cm]{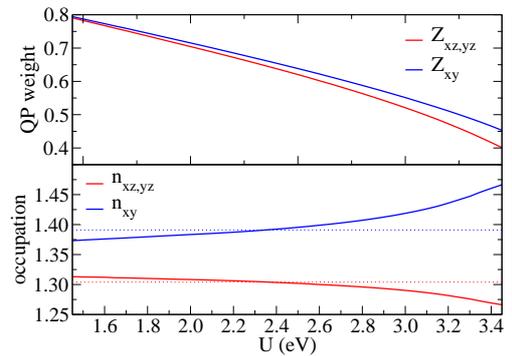}
\caption{Top: Orbital-dependent QP weight (top) and orbital
occupations per Ru ion (bottom) with increasing $U$ for the $t_{2g}$ states.
The dotted line in the bottom part marks the respective occupation for $U$=0.
\label{pic:qpocc}}
\end{center}
\end{figure}
According to theoretical estimates for the ruthenates~\cite{lie00,oka04,mra10}, 
the value of the Hund's coupling was set to $J$=0.35 eV for all the computations. 
While there are LDA+DMFT investigations for Sr$_2$RuO$_4$ with values for the 
Hubbard $U$ between 1-3 eV~\cite{lie00,pch07,mra10}, to our knowledge no such 
approach exists to Sr$_3$Ru$_2$O$_7$. Yet optics experiments~\cite{puc98} point 
towards a similar $U$ value. Here we scanned the onsite Coulomb interaction up to 
$U$=3.45 eV, since it is also known that due to the neglect of quantum fluctuations
in slave-boson approaches, the effect of $U$ may be underestimated~\cite{gri10}.

Figure~\ref{pic:qpocc} shows the diagonal QP weight $Z_{mm}$ for $m$=$(xz,yz),xy$
with respect to $U$. The inter-orbital terms $Z_{mm'}$ remain marginal and appear 
irrelevant in the present orbital representation. It is seen that moderate $U$ 
values already provide a significant QP renormalization. The latter is smaller for 
the $d_{xy}$ orbital, understandable from the larger orbital-resolved bandwidth. 
In addition, Fig.~\ref{pic:qpocc} displays the orbital- and $U$-dependent 
occupations for the total four electrons on each Ru ion. As explainable from the
lower crystal-field level, the $d_{xy}$ orbital is stronger occupied in the
noninteracting case. It may be observed that the affect of $U$ on the occupations
is rather subtle, with compensating/polarizing tendencies below/above 
$U^*$$\sim$2.35 eV. Note of course that the latter value depends strongly on the
chosen value for $J$. Though the correlation-induced interorbital charge transfers
are not dramatic, one may still expect impact on the very sensitive low-energy
physics of Sr$_3$Ru$_2$O$_7$.

\begin{figure}[t]%
\begin{center}
\includegraphics*[width=7.75cm]{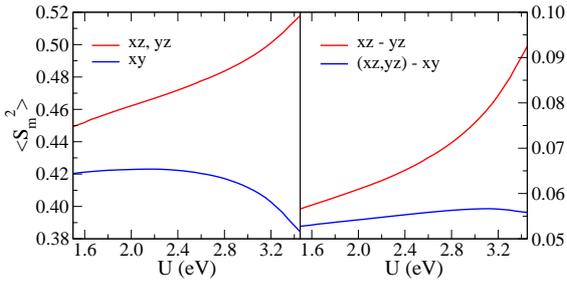}
\caption{Local intra- and inter-orbital spin correlations with increasing
$U$.
\label{pic:spincorr}}
\end{center}
\end{figure}

\begin{figure}[t]%
\begin{center}
\includegraphics*[width=5.75cm]{SBA.eps}\hspace*{0.2cm}
\includegraphics*[width=2cm]{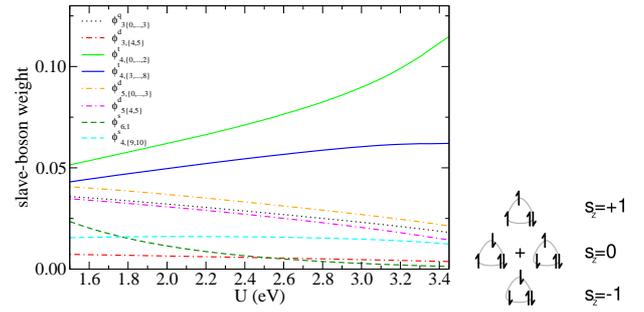}
\caption{Left: Weights of the local multiplets on a given Ru ion via the 
symmetry-adapted slave-boson amplitudes squared 
$|\phi^{\hfill}_{\Gamma\Gamma}|^2$ with 
$U$. Right: Fock decomposition of the dominant threefold degenerate triplet state 
$\Gamma_{4,\{0,\ldots,2\}}^{\rm t}$, where $d_{xy}$ is the doubly-occupied orbital. 
\label{pic:sba}}
\end{center}
\end{figure}

According to Hund's rule, in an atomic picture the Ru ion should have $S$=1 with
the $d_{xy}$ orbital being doubly occupied on the basis of the given crystal-field
splitting. Our
calculations indeed show that $S^2$=$S(S$$+$$1)$ increases from $S^2$$\sim$1 at 
$U$=0 towards $S^2$$\sim$1.83 for $U$=3.45 eV. In this respect, the 
orbital-resolved local spin correlations are shown in Fig.~\ref{pic:spincorr}, 
with the anticipated different behavior for $d_{xy}$ and $d_{xz,yz}$. Namely, 
$S_{xz,yz}^2$ grows strongly with $U$, whereas $S_{xy}^2$ diminishes substantially
after $U^*$ in order to cope with the formation of the local spin pair in that
orbital. Also the inter-orbital spin correlations show the designated $xz$-$yz$
spin-parallel coupling. Another option to investigate the local spin states is given
by an inspection of the slave-boson weights $|\phi_{\Gamma\Gamma'}|^2$ within the
local Ru($t_{2g}$) multiplet basis $\{\Gamma\}$. Therefore the slave-boson
amplitudes $\phi_{nn'}$ are rotated in the eigenbasis of the isolated local
Hamiltonian~(\ref{eq:locham}) via
\begin{equation}
\phi^{\hfill}_{\Gamma\Gamma'}=\mathcal{U}^\dagger_{\Gamma n}\phi^{\hfill}_{nn'}
\mathcal{U}^{\hfill}_{n'\Gamma'}\quad,
\end{equation}
where ${\cal U}_{n\Gamma}$ provides the unitary mapping between the multiplet basis
$\{\Gamma\}$ and the Fock basis $\{n\}$. We denote a specific multiplet by 
$\Gamma_{p,r}^{m}$, where $p$ describes the particle sector, $r$ the 
energy level therein (starting with $n$=0 for the respective ground state)
and $m$ marks the spin state, i.e., singlet 's', doublet 'd', triplet 't' and 
quartet 'q'. The diagonal multiplet weights $|\phi_{\Gamma\Gamma}|^2$ are 
plotted in Fig.~\ref{pic:sba} with respect to $U$. Note that in the present case
the $\phi^{\hfill}_{\Gamma\Gamma'}$ are close to diagonal with only few 
off-diagonal terms of minor amplitude. It is seen that the local physics is 
dominated by a threefold-degenerate triplet state in the four-particle sector
with $d_{xy}$ indeed being the doubly-occupied orbital. Beyond that one, 
two other triplets with only marginal energy difference follow and then already
a five-particle doublet shows up relevant, however loosing weight with increasing
$U$.

\begin{figure}[t]%
\begin{center}
\includegraphics*[width=8cm]{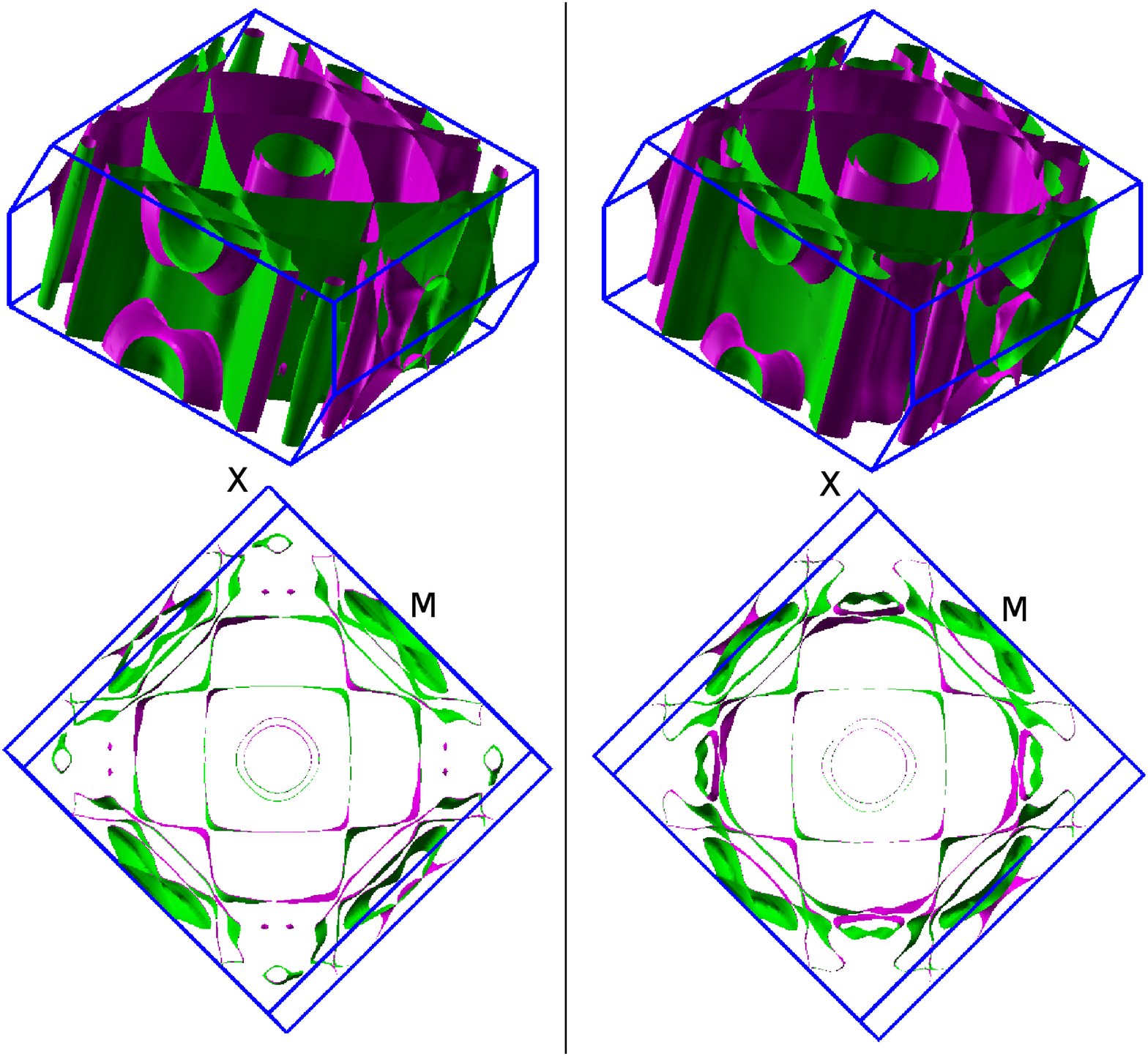}
\caption{Fermi surface of Sr$_3$Ru$_2$O$_7$ for $U$=0 (left) and $U$=3.45 eV
(right). In the bottom the view along the $c^*$ axis in k-space is depicted.
\label{pic:fermi}}
\end{center}
\begin{center}
\includegraphics*[width=7.75cm]{renormBANDS.eps}
\caption{LDA (dashed lines) and renormalized QP bands ($U$=3.45 eV) along the 
high-symmetry directions. The lower part shows a blow up around the Fermi level 
$\varepsilon_{\rm F}$ with the FS sheets  
$\delta$, $\alpha_1$, $\alpha_2$, $\gamma_1$, $\beta$ and $\gamma_2$.
\label{pic:renormb}}
\end{center}
\end{figure}

Besides affecting the local states, the electronic correlations introduced by
$U$ and $J$ of course also have impact on the QP dispersions. From 
eq.~(\ref{eq:Sigma_physical}) it follows that the RISB self-energy {\sl narrows} 
the bands via $Z$ and gives rise to band {\sl shifts} through the remaining static
term. In Fig.~\ref{pic:fermi} we compare the Fermi surface (FS) in LDA ($U$=0)
with the case for $U$=3.45 eV. In general the FS is rather complex with in total
six sheets, labeled $\delta$, $\alpha_1$, $\alpha_2$, $\gamma_1$, $\beta$ and
$\gamma_2$ (see e.g.~Ref.~\cite{tam08}). The LDA FS is in rather good agreement with
ARPES data~\cite{tam08}. The overall appearance bears the prominent 
quasi-twodimensional character of the compound with only weak FS warping along
$c^*$, especially for the sheets in the inner area of the BZ. However there are
also bands crossing the Fermi level along $c^*$ close to the zone boundary, e.g.
from $M$$-$$R$. This observation may be important for an understanding of the 
apparent metallic behavior of the measured optical conductivity along 
$c$~\cite{mir08}. The interacting FS shows for the chosen $U$ value no 
dramatic differences, which is expected since it is well known that LDA provides
a surprisingly good fermiology for many strongly correlated metals. However
there are still some important changes that may shed some further light on the 
generic low-energy physics of Sr$_3$Ru$_2$O$_7$. The renormalized FS can partly
also be studied from the deviations of the associated bands crossing 
$\varepsilon_{\rm F}$ along the high-symmetry lines in the BZ, as displayed in 
Fig.~\ref{pic:renormb}.
For most of the sheets there are some minor size
changes with $U$ (which should be in overall accordance with Luttinger's theorem),
e.g., $\delta$, $\alpha_{1}$ and $\alpha_2$ somewhat growing in the $k_z$=0 plane.
However the $\gamma_2$ pocket is rather severely modified. First of all from ARPES 
experiment, $\gamma_2$ appears as a small but ``simple'' hole pocket more or less 
right inbetween the $\alpha_2$ sheet boundary and the $X$ point along the 
$\Gamma$$-$$X$ direction~\cite{tam08}. On the other hand, the computed (QP) band 
structure in this region looks rather complex, with possibly as many as four 
Fermi-level crossings. Intriguingly, those crossings pair into two with two bands 
crossing each other in an ``dirac-cone-manner'', respectively (cf. 
Fig.~\ref{pic:renormb}).
As it turns out, this low-energy structure is very sensitive to electronic 
correlations and depending on the interaction strength, the two named crossing
twofolds shift between hole- or electron-like fermiology. In the LDA case the
twofold close to $X$ is higher in energy, giving rise to a hole pocket as 
observed in experiment, whereas the twofold close to $\alpha_2$ is barely
shifted into the electron-pocket appearance. For $U$=3.45 eV the situation is
nearly reversed for the former pocket part, while the latter becomes stabilized
in whats now a sizeable electron pocket. Note that since being close to the 
$\gamma_2$-pocket structure also the $\alpha_2$ sheet in the $\Gamma$$-$$X$ 
direction is affected by the low-energy correlation-induced restructering. This
results in a stronger renormalization of the $\alpha_2$ sheet in that direction
compared to the one along $\Gamma$$-$$M$, which is in agreement with findings in
recent ARPES experiments~\cite{lee09}.

\begin{figure}[b]%
\begin{center}
\includegraphics*[width=7.75cm]{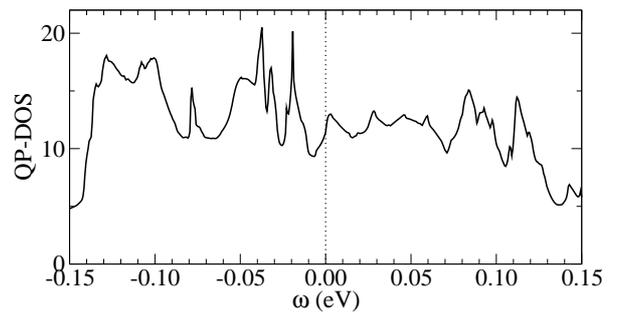}
\caption{RISB QP density of states for the $t_{2g}$ states per Ru ion 
($U$=3.45 eV). Note that 
the absolute value is somewhat artificial due to the not-described Hubbard bands 
in the spectral function within slave-boson mean-field and should not be directly
related to experimental values.
\label{pic:renormd}}
\end{center}
\end{figure}
Surely, this whole discussion depends on the magntiude of $U$ and $J$, however 
already the fact that underneath the $\gamma_2$ pocket hides such a 
correlation-sensitive low-energy structure may be of greater importance. The reason
for the sensitivity is besides the flatness of the associated bands close to the
$X$ point given by the fact that while the two hole-like bands of the outlined
fourfold are dominantly $d_{xy}$, the electron-like bands are mainly $d_{xz,yz}$. 
Thus the multi-orbital character of the compound directly manifests in the complex
structure around the $\gamma_2$ pocket. In the experimental 
dispersions~\cite{tam08} the described low-energy $\gamma_2$ structure is partly 
visible in the occupied area, however much more strongly renormalized 
(lowest-energy band $\sim$1 meV). The effect of SOC, especially via additional band 
splittings, may here be important to reveal more details of this low-energy 
segment. In addition, the renormalization effect through quantum spin fluctuations
could be relevant in the system since its prone to the ferromagnetic instability.

Finally, we show in Fig.~\ref{pic:renormd} the QP DOS from the LDA+RISB calculation
in a small energy window around $\varepsilon_{\rm F}$.
Albeit we obviously can not fully reach in the present modeling the very 
low-energy scale seen in experiment, it is observable that already here a rather
subtle energy structure shows up in a small range around 
$\varepsilon_{\rm F}$. The latter value resides for $U$=3.45 eV within a slope of
width $\sim$10 meV and onset 6 meV below zero energy. Note that a large number of 
unoccupied states above $\varepsilon_{\rm F}$ is also verified from ARPES 
measurements~\cite{tam08}.

\section{Conclusions}
The present work dealt with a realistic LDA+RISB approach to the low-energy 
electronic structure of the puzzling Sr$_3$Ru$_2$O$_7$ compound. It was shown
that the LDA band structrure close to the Fermi level can be accurately 
downfolded to an effective MLWF description with tailored $t_{2g}$-like orbitals.
Supplemented by generic on-site Coulomb correlations, the resulting interacting
model accounts for strong electronic self-energy effects. A dominant filling 
of the $d_{xy}$ orbital was identified, locally favoring a corresponding 
four-particle triplet state at larger Hubbard $U$. The renormalization of the
low-energy electronic structure singles out the $\gamma_2$ pocket 
(and its neighborhood) as especially sensitive to electronic correlations, also
due to its manifest multi-orbital character. In view of the very low energy scale
close to the $X$ point this observation draws much attention to that pocket as
a possible key object for the understanding of metamagnetism in 
Sr$_3$Ru$_2$O$_7$~\cite{pue10}.
Albeit the Hubbard parameters have been chosen more or less by hand in this study,
the mere fact that one may here find definite regions in {\bf k}-space that are 
rather prominently susceptible to electronic correlations seems an interesting
result that deserves further detailed investigation. The neglect of SOC is a
sure drawback of the present study, but because of the already good data agreement
between experiment and theory on, e.g., the Fermi surface, very strong changes of
the current results are not expected with including SOC effects. The latter will
be of course important when turning on a magnetic field H, especially when
elaborating on the influence of the angle-dependence thereof. However the RISB 
formalism is in principle ideally suited to cope with such physics and we plan to 
address it in a future work. For instance, the orbital-occupation shifts in the 
correlated regime with respect to an angle-dependent field, possibly close to the 
metamagnetic response, would be of vital interest.

\begin{acknowledgement}
Financial support from the Free and Hanseatic City of Hamburg in the context of 
the NANOSPINTRONICS Landesexzellenzinitiative is gratefully acknowledged. Computations
were performed at the North-German Supercomputing Alliance (HLRN).
\end{acknowledgement}

%
\bibliographystyle{pss}
\bibliography{bibextra}

\providecommand{\WileyBibTextsc}{}
\let\textsc\WileyBibTextsc
\providecommand{\othercit}{}
\providecommand{\jr}[1]{#1}
\providecommand{\etal}{~et~al.}


\begin{thebibliography}{[10]}

\bibitem{col84}
 \textsc{P.~Coleman},
 \jr{Phys. Rev. B} \textbf{29}, 3035 (1984).


\bibitem{kot86}
 \textsc{G.~Kotliar} and  \textsc{A.\,E. Ruckenstein},
 \jr{Phys. Rev. Lett.} \textbf{57}, 1362 (1986).


\bibitem{li89}
 \textsc{T.~Li},  \textsc{P.~W\"olfle},  and  \textsc{P.\,J. Hirschfeld},
 \jr{Phys. Rev. B} \textbf{40}, 6817 (1989).


\bibitem{lec07}
 \textsc{F.~Lechermann},  \textsc{A.~Georges},  \textsc{G.~Kotliar},  and
  \textsc{O.~Parcollet},
 \jr{Phys. Rev. B} \textbf{76}, 155102 (2007).


\bibitem{gri01}
 \textsc{S.\,A. Grigera},  \textsc{R.\,S. Perry},  \textsc{A.\,J. Schofield},
  \textsc{M.~Chiao},  \textsc{S.\,R. Julian},  \textsc{G.\,G. Lonzarich},
  \textsc{S.\,I. Ikeda},  \textsc{Y.~Maeno},  \textsc{A.\,J. Millis},  and
  \textsc{A.\,P. Mackenzie},
 \jr{Science} \textbf{294}, 329 (2001).


\bibitem{geg06}
 \textsc{P.~Gegenwart},  \textsc{F.~Weickert},  \textsc{M.~Garst},
  \textsc{R.\,S. Perry},  and  \textsc{Y.~Maeno},
 \jr{Phys. Rev. Lett.} \textbf{96}, 136402 (2006).


\bibitem{bor06}
 \textsc{R.\,A. Borzi},  \textsc{S.\,A. Grigera},  \textsc{J.~Farrell},
  \textsc{R.\,S. Perry},  \textsc{S.\,J.\,S. Lister},  \textsc{S.\,L. Lee},
  \textsc{D.\,A. Tennant},  \textsc{Y.~Maeno},  and  \textsc{A.\,P. Mackenzie},
 \jr{Science} \textbf{315}, 214 (2006).


\bibitem{mac03}
 \textsc{A.\,P. Mackenzie} and  \textsc{Y.~Maeno},
 \jr{Rev. Mod. Phys.} \textbf{75}, 657 (2003).


\bibitem{sha00}
 \textsc{H.~Shaked},  \textsc{J.\,D. Jorgensen},  \textsc{O.~Chmaissem},
  \textsc{S.~Ikeda},  and  \textsc{Y.~Maeno},
 \jr{J. Solid State Chem.} \textbf{154}, 361 (2000).


\bibitem{ike00}
 \textsc{S.\,I. Ikeda},  \textsc{Y.~Maeno},  \textsc{S.~Nakatsuji},
  \textsc{M.~Kosaka},  and  \textsc{Y.~Uwatoko},
 \jr{Phys. Rev. B} \textbf{62}, R6089 (2000).


\bibitem{perr01}
 \textsc{R.\,S. Perry},  \textsc{L.\,M. Galvin},  \textsc{S.\,A. Grigera},
  \textsc{L.~Capogna},  \textsc{A.\,J. Schofield},  \textsc{A.\,P. Mackenzie},
  \textsc{M.~Chiao},  \textsc{S.\,R. Julian},  \textsc{S.\,I. Ikeda},
  \textsc{S.~Nakatsuji},  and  \textsc{Y.~Maeno},
 \jr{Phys. Rev. Lett.} \textbf{86}, 2661 (2001).


\bibitem{gri03}
 \textsc{S.\,A. Grigera},  \textsc{R.\,A. Borzi},  \textsc{A.\,P. Mackenzie},
  \textsc{S.\,R. Julian},  \textsc{R.\,S. Perry},  and  \textsc{Y.\,M. and},
 \jr{Phys. Rev. B} \textbf{67}, 214427 (2003).


\othercit
\bibitem{mbpp_code}
 \textsc{B.~Meyer},  \textsc{C.~Els\"{a}sser},  \textsc{F.~Lechermann},  and
  \textsc{M.~F\"{a}hnle},
FORTRAN 90 Program for Mixed-Basis-Pseudopotential Calculations for Crystals,
Max-Planck-Institut f\"{u}r Metallforschung, Stuttgart, unpublished.


\bibitem{lou79}
 \textsc{S.\,G. Louie},  \textsc{K.\,M. Ho},  and  \textsc{M.\,L. Cohen},
 \jr{Phys. Rev. B} \textbf{19}, 1774 (1979).


\bibitem{van85}
 \textsc{D.~Vanderbilt},
 \jr{Phys. Rev. B} \textbf{32}, 8412 (1985).


\bibitem{per92}
 \textsc{J.\,P. Perdew} and  \textsc{Y.~Wang},
 \jr{Phys. Rev. B} \textbf{45}, 13244 (1992).


\bibitem{mar97}
 \textsc{N.~Marzari} and  \textsc{D.~Vanderbilt},
 \jr{Phys. Rev. B} \textbf{56}, 12847 (1997).


\bibitem{sou01}
 \textsc{I.~Souza},  \textsc{N.~Marzari},  and  \textsc{D.~Vanderbilt},
 \jr{Phys. Rev. B} \textbf{65}, 035109 (2001).


\bibitem{has97}
 \textsc{I.~Hase} and  \textsc{Y.~Nishihara},
 \jr{J. Phys. Soc. Jpn.} \textbf{66}, 3517 (1997).


\bibitem{sin01}
 \textsc{D.~Singh} and  \textsc{I.\,I. Mazin},
 \jr{Phys. Rev. B} \textbf{63}, 165101 (2001).


\bibitem{tam08}
 \textsc{A.~Tamai},  \textsc{M.\,P. Allan},  \textsc{J.\,F. Mercure},
  \textsc{W.~Meevasana},  \textsc{R.~Dunkel},  \textsc{D.\,H. Lu},
  \textsc{R.\,S. Perry},  \textsc{A.\,P. Mackenzie},  \textsc{D.\,J. Singh},
  \textsc{Z.\,X. Shen},  and  \textsc{F.~Baumberger},
 \jr{Phys. Rev. Lett.} \textbf{101}, 026407 (2008).


\bibitem{lee09}
 \textsc{J.~Lee},  \textsc{M.\,P. Allan},  \textsc{M.\,A. Wang},
  \textsc{J.~Farrell},  \textsc{S.\,A. Grigera},  \textsc{F.~Baumberger},
  \textsc{J.\,C. Davis},  and  \textsc{A.\,P. Mackenzie},
 \jr{Nat. Phys.} \textbf{5}, 800 (2009).


\bibitem{lie00}
 \textsc{A.~Liebsch} and  \textsc{A.~Lichtenstein},
 \jr{Phys. Rev. Lett.} \textbf{84}, 1591 (2000).


\bibitem{oka04}
 \textsc{S.~Okamoto} and  \textsc{A.\,J. Millis},
 \jr{Phys. Rev. B} \textbf{70}, 195120 (2004).


\bibitem{mra10}
 \textsc{J.~Mravlje},  \textsc{M.~Aichhorn},  \textsc{T.~Miyake},
  \textsc{K.~Haule},  \textsc{G.~Kotliar},  and  \textsc{A.~Georges},
 \jr{arXiv:1010.5910v1} (2010).


\bibitem{pch07}
 \textsc{Z.\,V. Pchelkina},  \textsc{I.\,A. Nekrasov},  \textsc{T.~Pruschke},
  \textsc{A.~Sekiyama},  \textsc{S.~Suga},  \textsc{V.\,I. Anisimov},  and
  \textsc{D.~Vollhardt},
 \jr{Phys. Rev. B} \textbf{75}, 035122 (2007).


\bibitem{puc98}
 \textsc{A.\,V. Puchkov},  \textsc{M.\,C. Schabel},  \textsc{D.\,N. Basov},
  \textsc{T.\,S. abd G.~Cao},  \textsc{T.~Timusk},  and  \textsc{Z.\,X. Shen1},
 \jr{Phys. Rev. Lett.} \textbf{81}, 2747 (1998).


\bibitem{gri10}
 \textsc{D.~Grieger},  \textsc{L.~Boehnke},  and  \textsc{F.~Lechermann},
 \jr{J. Phys.: Condens. Matter} \textbf{22}, 275601 (2010).


\bibitem{mir08}
 \textsc{C.~Mirri},  \textsc{L.~Baldassarre},  \textsc{S.~Lupi},
  \textsc{M.~Ortolani},  \textsc{R.~Fittipaldi},  \textsc{A.~Vecchione},  and
  \textsc{P.~Calvani},
 \jr{Phys. Rev. B} \textbf{78}, 155132 (2008).


\bibitem{pue10}
 \textsc{C.\,M. Puetter},  \textsc{J.\,G. Rau},  and  \textsc{H.\,Y. Kee},
 \jr{Phys. Rev. B} \textbf{81}, 081105(R) (2010).


\end{thebibliography}
%


\end{document}